\documentstyle[12pt]{article}
\begin{document}

\def\e{{\rm {e}}}
\def\ba{\begin{array}}
\def\ea{\end{array}}
\def\bea{\begin{eqnarray}}
\def\eea{\end{eqnarray}}
\def\bea*{\begin{eqnarray*}}
\def\eea*{\end{eqnarray*}}
\def\be{\begin{equation}}
\def\ee{\end{equation}}

\begin{center}

{\large \bf SCATTERING IN THE PAIR OF HILBERT SPACES}

\bigskip
{\bf E.A.~Bugrij,
A.V.~Mishchenko\footnote{{\it e-mail:} olmishch@ap3.bitp.kiev.ua},
Yu.A.~Sitenko\footnote{{\it e-mail:} yusitenko@bitp.kiev.ua}}

\medskip
{Bogolyubov Institute for Theoretical Physics}

\medskip
{252142, Kiev-142, Ukraine}

\end{center}

\bigskip
\begin{abstract}
\begin{sloppypar}
The problem of the scattering of a charged test particle in the
gravitational background of axially symmetrical wormhole
in the presence of the Aharonov-Bohm type magnetic field is considered.
It is shown that the natural mathematical framework appropriate
for the problem is the scattering theory in the pair of Hilbert spaces.
Both relevant modified wave operators and $S$-matrices are found.
\end{sloppypar}
\end{abstract}

\bigskip
\bigskip

Let us consider the problem of the scattering of a charged test particle in
the cylindrically-symmetric curved space with the topology of (cylinder
$\times R^2$) in the presence of the static magnetic field of Aharonov-Bohm
type in the sense described below. At first, we restrict ourselves by
considering static cylindrically-symmetric configuration of a gravitational
field with metric
\be
ds^2=dt^2-dz^2 - f(r)\left[dr^2 + r^2 d\theta^2\right]
\ee
where $r$, $\theta$ are polar coordinates in the transverse $z$ axis
direction, while $z$ is a coordinate in the direction of the space
symmetry axis. For $f(r)$ only its conical behavior for $r\to \infty$
and $r\to 0$:
$$
f(r)=\biggl(\frac{r}{R_1}\biggr)^{-2\Phi_k^{(1)}}, \quad r\geq R_1,
\quad  \Phi_k^{(1)}<1,
$$
\be
f(r)=\biggl(\frac{r}{R_2}\biggr)^{-2\Phi_k^{(2)}}, \quad r\leq R_2,
\quad  \Phi_k^{(2)}>1,
\ee
as well as continuity condition, $f(r)\in C(0,\infty)$, are
assumed. In (2), $R_1$ and $R_2$ are arbitrary constants,
$0<R_1$, $R_2<\infty$.  The
above-mentioned gravitational structure is, in fact, a classical
static wormhole, i.e. the spatial configuration consisting of two
noncompact asymptotically flat conic regions connected by a compact
strongly curved region -- wormhole throat.

It is well known that the Aharonov-Bohm type field strength is zero in the
region reachable for particles. The latter can be provided whether by
puncturing the point in which the field strength is infinite (for the
classical Aharonov-Bohm field) or by restriction of the particle motion  to
the region by means of potential barrier, for example. In our case, we can
use pure geometrical device that disjoitns the region with non-zero magnetic
field strength from charged matter. Namely, we can imagine that this region
is concealed inside wormhole.

Motion of a test particle in the
exterior along the wormhole axis is free, therefore, our concern will be on
the $z={\rm const}$ section. The dynamics of the test particle
under consideration is governed by the Hamiltonian:
\be
H=\frac{\hbar^2}{2m}\biggl[-\partial^2_r-\frac{1}{r}\partial_r+
\frac{1}{r^2}\biggl(i\partial_\theta+\Phi+
\frac{\sigma^3}{4}r\partial_r\ln f\biggr)^2-
\frac{1}{4}\biggl(\partial^2_r\ln f+ \frac{1}{r}\partial_r\ln
f\biggr)\biggr],
\ee
where
$$
\sigma^3=\left(
\begin{array}{cc}
1 & 0 \\
0 & -1
\end{array}
\right)\;,
$$
in the case of a spinor particle ($\psi$ is a two-component column
function) and $\sigma = 0$ in the case of a scalar particle
($\psi$ is a single-component function),
$0\leq \Phi<1$ is
magnetic flux and scalar product in the Hilbert  space ${\cal {H}}$ of
square integrable functions
on the curved space $M$ is following:
\be
(\psi,\psi')=\int\limits_0^{2\pi}d\theta\int\limits_0^\infty dr\,r\,
f(r)\psi^\dagger \psi',
\ee
where sign $\dagger$ denotes hermitian conjugation.

It is known that to formulate the problem of scattering it is necessary, at
least, to determine two Hamiltonians, $H_0$ and $H$. The first one describes
non-perturbed dynamics and the second one -- perturbed dynamics. Having
determined the above-mentioned pair of Hamiltonians one builds wave
(M\"oller) operators
\be
\Omega^{\pm}(H,H_0)=s-\lim_{t\to\pm\infty}\e^{iHt}\e^{-iH_0t}P(H_0).
\ee
where $P(H_0)$ is a projector on absolutely continuous subspace of
operator $H_0$.
Let us note that expression (5) makes sense only in the case, when
perturbation
\be
V=H-H_0
\ee
has a short range (see \cite{1.}, \cite{2.}). For the second
order operators the latter means
(following H\"ormander) \cite{2.} that the coefficients in the expansion
\be
V=a_{ij}\partial_i\partial_j+b_i\partial_i+c
\ee
decrease like
$|r|^{-1-\varepsilon}$
 for
$|r|\to\infty$
  and
$\varepsilon>0$.
   For the case (following H\"ormander)
when perturbation has long range, when the above-mentioned (real)
coefficients decrease like
$|r|^{-\varepsilon}$
for
$|r|\to \infty$, $\varepsilon>0$
 the definition of wave
operators (5) must be modified. At last, we cannot say anything as to the
existence of the wave operators in the case when H\"ormander condition of
long range action is broken.

We want to consider scattering of a particle
which dynamics coincides with one of a free particle on the cone in the
infinitely remote past. As it is known \cite{3.}, the latter is described
simply by
free Hamiltonian on the plane $R^2$
\be
H_0=-{\hbar^2\over2m}\Delta_2.
\ee
This is conditioned by the fact that cone is a flat surface with zero Gauss
curvature everywhere except apex and, consequently, locally coincides with
$R^2$. Analogously, it is supposed that in the infinitely remote future
reflected wave will exist. Its dynamics is determined by Hamiltonian $H_0$.
The wave which has passed through the wormhole throat will exist, too. And
its dynamics is determined by Hamiltonian $H_0$, but acting on
functions that
are defined on the second cone. It turns out, that in such formulation of
scattering problem it is impossible to choose Hamiltonian describing a free
dynamics in all curved space $M$.

Nevertheless, mathematical formalism appropriate for our problem exists, the
scattering theory in a pair of Hilbert spaces, namely \cite{1.}.

In the problem it is more convenient to use a new radial coordinate
\be
\rho(r)=\int\limits^r_{r_0}dr'[f(r')]^{1/2}
\ee
instead of $r$, where $r_0$ is arbitrary parameter. The metric (1) then has
the form:
\be
ds^2=d\rho^2+g_{00}(\rho)d\theta^2\;.
\ee
Let us perform the unitary mapping
\be
{\cal{H}} \to \widetilde{\cal{H}}, \qquad
\psi \to \widetilde{\psi} = r^{1/2} f^{1/4} \psi
\ee
where
$\widetilde{\cal{H}}$
 is Hilbert space with scalar product
\be
(\widetilde{\psi},\widetilde{\psi}')=\int\limits_0^{2\pi}d\theta
\int\limits^\infty_{-\infty}d\rho\,\widetilde{\psi}^\dagger
\widetilde{\psi}'.
\ee
Then the Hamiltonians $H$ and $H_0$ map into
\be
\widetilde{H}={\hbar^2\over2m}\biggl[-\partial^2_\rho+g_{00}^{-1}
\biggl(i\partial_\theta+\phi-{\sigma^3\over2}+{\sigma^3\over4}g_{00}^{1/2}
\partial_\rho\ln g_{00}\biggr)^2-
{1\over16}\biggl(\partial_\rho\ln g_{00}\biggr)^2\biggr],
\ee
\be
\widetilde{H}_0=-{\hbar^2\over2m}\biggl[\partial^2_\rho+
{\partial^2_{\theta}\over\rho^2}+{1\over4\rho^2}\biggr].
\ee
In the further consideration we will omit tilde sign.

In addition to the Hilbert space
${\cal H}$ let us define two Hilbert spaces $\cal{H}_+$ and
$\cal{H}_-$. Space $\cal{H}_+$ is a space of square integrable
functions on the right cone
$\theta\in[0,2\pi]$, $\rho\in [0,+\infty)$
 with a scalar product
\be
(\psi,\psi')=\int\limits_0^{2\pi}d\theta\int\limits_0^\infty
d\rho\psi^\dagger \psi'.
\ee
Analogously, the space $\cal{H}_-$ is a space of square integrable
functions on the left cone
$\theta\in[0,2\pi]$, $\rho\in [-\infty,0)$
 with a scalar product
\be
(\psi,\psi')=\int\limits_0^{2\pi}d\theta\int\limits_{-\infty}^0
d\rho\psi^\dagger \psi'.
\ee
 Let us
also define operators
$J_{\pm}$
 which map the Hilbert spaces
$\cal{H}_{\pm}$
  into
$\cal{H}$
   and
act as follows. For
$\psi_{\pm}\in\cal{H}_{\pm}$:
\be
J_{\pm}\psi_{\pm}=\mu_{\pm}(\rho)\psi_{\pm},
\ee
where
$\mu_{\pm}(\rho)\in C^1(\pm\infty,0)$
 and
$$
\mu_{\pm}(\rho)=0, \qquad |\rho|<\varepsilon,
$$
\be
\mu_{\pm}(\rho)=1, \qquad |\rho|>\cal{P},
\ee
where
$0<\varepsilon<\cal{P}<\infty$.
 To solve the problem of scattering formulated earlier it is
necessary to build the solution of Schr\"odinger equation
\be
i\hbar{\partial\psi\over\partial t}=H\psi,
\ee
that satisfies in the infinitely remote past the condition
\be
\lim_{t\to-\infty}||\psi(t)-J_+\e^{-iH_0^{(+)}t}\phi_+^{(-)}||=0
\ee
and in the infinitely remote future the conditions
$$
\lim_{t\to+\infty}||\psi(t)-J_+\e^{-iH_0^{(+)}t}\phi_+^{(+)}||=0,
$$
\be
\lim_{t\to+\infty}||\psi(t)-J_-\e^{-iH_0^{(-)}t}\phi_-^{(+)}||=0.
\ee
Here,
$\phi_+^{(\pm)}\in\cal{H}_+$
 and
$\phi_-^{(+)}\in\cal{H}_-$,
the Hamiltonians
$H_0^{(+)}$
  and
$H_0^{(-)}$
   describe a free dynamics on
the right and left cones, respectively, and in compliance with (14) have
the form:
\be
H_0^{(\pm)}=-
{\hbar^2\over2m}\biggl[\partial^2_\rho+{\partial^2_0\over\rho^2}+
{1\over4\rho^2}\biggr].
\ee
Two scattering matrices
$S_{++}$
 and
$S_{+-}$
\be
\phi_+^{(+)}=S_{++}\phi_+^{(-)}, \qquad
\phi_-^{(+)}=S_{+-}\phi_+^{(-)}
\ee
will contain all necessary information about scattering. Taking into account
that
$\psi(t)=\e^{-iHt}\psi_0$
for some
$\psi_0\in\cal{H}$,
 we get
\begin{eqnarray}
S_{++} & = & \bigl(\Omega_+^{(-)}(H,H_0^{(+)};J_+)\bigr)^{\dagger}
\Omega_+^{(+)}(H,H_0^{(+)};J_+),\nonumber\\
S_{+-} & = & \bigl(\Omega_-^{(-)}(H,H_0^{(-)};J_-)\bigr)^{\dagger}
\Omega_+^{(+)}(H,H_0^{(+)};J_+),
\end{eqnarray}
where wave operators
$\Omega_+^{(+)}$, $\Omega_+^{(-)}$, $\Omega_-^{(-)}$
 are determined by expressions:
$$
\Omega_+^{(+)}(H,H_0^{(+)};J_+)=s-\lim_{t\to-\infty}\e^{iHt}J_+
\e^{-iH_0^{(+)}t},
$$
$$
\Omega_+^{(-)}(H,H_0^{(+)};J_+)=s-\lim_{t\to+\infty}\e^{iHt}J_+
\e^{-iH_0^{(+)}t},
$$
\be
\Omega_-^{(-)}(H,H_0^{(-)};J_-)=s-\lim_{t\to+\infty}\e^{iHt}J_
-\e^{-iH_0^{(-)}t}.
\ee
In the expressions (25) we take into account that the spectra of operators
$H_0^{(+)}$
and
$H_0^{(-)}$
 are absolutely continuous.

Thus, the problem of scattering is reduced to the problem of finding wave
operators
$\Omega_+^{(\pm)}$
 for the pair of Hilbert spaces
$\cal{H}_+$
 and
$\cal{H}$,
  and wave operator
$\Omega_-^{(-)}$
for the pair of Hilbert spaces
$\cal{ H}_-$ and   $\cal{ H}$.
  Let us construct them. One can use
the methods of stationary scattering theory and find wave functions
$\psi_+^{(\pm)}(\vec{k})$
 and
$\psi_-^{(-)}(\vec{k})$
instead of wave operators. These functions are the result of action of wave
operators on functions
\be
\psi_{\pm}^{(0)}(k)={1\over2\pi}\e^{ik\rho\cos(\theta-\theta_k)}\;,
\ee
where
$\theta_k$
 is the angle that determines the direction of wave vector
$\vec{k}$
  and
$k=|\vec{k}|$; $\rho\in(-\infty,0)$
 for
$\psi_-^{(0)}$
  and
$\rho\in(0,+\infty)$
   for
$\psi_+^{(0)}$
and the functions
$\psi_{\pm}^{(0)}$
     are free stationary waves on the
right and left cones. Then
\be
\psi_{+}^{(\pm)}(\vec{k})=\Omega_{+}^{(\pm)}\psi_{\pm}^{(0)}(\vec{k}),\qquad
\psi_{-}^{(\pm)}(\vec{k})=\Omega_{-}^{(-)}\psi_{-}^{(0)}(\vec{k}).
\ee
While expansion on functions
$\psi_{\pm}^{(0)}(\vec{k})$
 is, in fact, Furier transformation for spaces
$\cal{H}_{\pm}$,
 then functions
$\psi_{+}^{(\pm)}(\vec{k})$
 and
$\psi_{-}^{(-)}(\vec{k})$
  completely define the wave operators
$\Omega_{+}^{(\pm)}$
   and
$\Omega_{-}^{(-)}$.
Using expressions (25) and (26), (27), and the fact that
\be
H_{0}^{(\pm)}\psi_{\pm}^{(0)}={\hbar^2k^2\over2m}
\ee
as well, one can simply get the following Lippmann-Shwinger equations:
$$
\psi_{+}^{(\pm)}(\vec{k})=J_+\psi_+^{(0)}(\vec{k})-G^{\pm}(\vec{k})
[HJ_+-J_+H_0^{(+)}]\psi_+^{(0)}(\vec{k}),
$$
\be
\psi_{-}^{(-)}(\vec{k})=J_-\psi_-^{(0)}(\vec{k})-G^{-}(\vec{k})
[HJ_--J_-H_0^{(-)}]\psi_-^{(0)}(\vec{k}),
\ee
where Green functions
\be
G^{\pm}(\vec{k})=\lim_{\varepsilon\to+0}\biggl[H-{\hbar^2k^2\over2m}\mp
i\varepsilon\biggr]^{-1}.
\ee

Taking into account the cylindrical symmetry of the problem, let us
   expend the functions $\psi_{+}^{(\pm)} $ and $\psi_{-}^{(-)}$ in
   the Furier series:
\be
\psi_{+}^{(\pm)}=\sum_{n=-
\infty}^{+\infty}\e^{in\theta}\psi_{+,n}^{(\pm)},\qquad
\psi_{-}^{(-)}=\sum_{n=-\infty}^{+\infty}\e^{in\theta}\psi_{-,n}^{(-)}.
\ee
The Hilbert space ${\cal{H}}$ will expand, correspondingly, to the
   direct sum
${\cal{H}}=\bigoplus\limits^{n=+\infty}_{n=-\infty}{\cal{H}}_n$, and
scalar product on
 ${\cal{H}}_n$ is defined as follows:
$$
(\psi_n,\psi'_n)=\int\limits_{-\infty}^{+\infty}d\rho\psi^\dagger_n\psi'_n.
$$

To construct the Green functions it is necessary to calculate
the resolvent of operator $H$:
\be
R_z=(H-z)^{-1}.
\ee
Let us choose the branch of
$\sqrt{z}$       that
$\Im \sqrt{z}\geq0$ and fix that
$\omega={\hbar\over\sqrt{2m}}\sqrt{z}$.  Then it is easy to get the
following expression for the kernel of the revolvent of operator $H$:
\be
R_z(\rho,\rho',\theta,\theta')={1\over2\pi}\sum_{n=-\infty}^{+\infty}
\e^{in(\theta-\theta')}R_n(\rho,\rho'),
\ee
where
\be
R_n(\rho,\rho')={i2m\over2a_n(\omega)\omega\hbar^2}\biggl[
\psi^{(n)}_1(\rho,\omega)
\psi^{(n)}_2(\rho',\omega)\theta(\rho-\rho')+
\psi^{(n)}_2(\rho,\omega)
\psi^{(n)}_1(\rho',\omega)\theta(\rho'-\rho)\biggr].
\ee
In expression (34)
$$
\theta(x)=\left\{
\begin{array}{l}
1\, ,\quad x > 0 \\
0\, ,\quad x < 0
\end{array}
\right.\
$$
and
$$
a_n(\omega)={i\over2\omega}W\{\psi^{(n)}_1,\psi^{(n)}_2\},
$$
where for two functions $f_1(\rho)$ and $f_2(\rho)$ Wronskian is defined
as follows
$$
W\{f_1(\rho),f_2(\rho)\} = f_1(\rho)\partial_{\rho}f_2(\rho) -
f_2(\rho)\partial_{\rho}f_1(\rho).
$$
The functions
$\psi^{(n)}_1(\rho,\omega)$ and
$\psi^{(n)}_2(\rho,\omega)$    are the pair of linearly independent
Jost solutions for partial Hamiltonian
$H_n=\e^{-in\theta}H\e^{in\theta}$,
which are characterized by following asymptotics:
$$
\psi^{(n)}_1(\rho,\omega)=\e^{i\omega\rho}, \qquad \rho\to+\infty,
$$
\be
\psi^{(n)}_2(\rho,\omega)=\e^{-i\omega\rho}, \qquad \rho\to-\infty.
\ee
Having defined the second pair of linearly independent Jost solutions
$$
\bar{\psi}^{(n)}_1(\rho,\omega)=\e^{-i\omega\rho}, \qquad
\rho\to+\infty,
$$
\be
\bar{\psi}^{(n)}_2(\rho,\omega)=\e^{i\omega\rho}, \qquad
\rho\to-\infty,
\ee
one can write the following useful expressions:
\be
\psi^{(n)}_1=a_n(\omega)\bar{\psi}^{(n)}_2+b_n(\omega)\psi^{(n)}_2,\quad
\psi^{(n)}_2=a_n(\omega)\bar{\psi}^{(n)}_1-\bar{b}_n(\omega)\psi^{(n)}_1;
\ee

\be
\bar{b}_n(\omega)=(b_n(\omega))^*=b_n(-\omega^*), \quad
(a_n(\omega))^*=a_n(-\omega^*);
\ee

\be
\bar{\psi}_i^{(n)}(\rho,\omega)=\psi_i^{(n)}(\rho,-\omega)=
(\psi_i^{(n)}(\rho,\omega^*))^*,\quad i=1,2;
\ee

$$
W\bigl\{\psi^{(n)}_1(\rho,\omega),\bar{\psi}^{(n)}_2(\rho,\omega)\bigr\}
=2ib_n(\omega)\omega,
$$
$$
W\bigl\{\psi^{(n)}_1(\rho,\omega),{\psi}^{(n)}_2(\rho,\omega)\bigr\}
=-2ia_n(\omega)\omega,
$$
\be
W\bigl\{\psi^{(n)}_2(\rho,\omega),\bar{\psi}^{(n)}_2(\rho,\omega)\bigr\}=
-W\bigl\{\psi^{(n)}_1(\rho,\omega),\bar{\psi}^{(n)}_1(\rho,\omega)\bigr\}
=2i\omega.
\ee
For real
$\omega$ let us fix
$k=\omega$.

Then the following expression takes place:
\be
|a_n(k)|^2-|b_n(k)|^2=1.
\ee
For the partial reflection and transmission coefficients one can write
\be
r_n(k)=b_n(k)/a_n(k),\qquad t_n(k)=1/a_n(k).
\ee
Then
\be
|r_n(k)|^2+|t_n(k)|^2=1
\ee
is partial unitarity condition.

The following scalar products
$$
\bigl(\psi_1^{(n)}(k),\psi_1^{(n)}(k')\bigr)=
2\pi |a_n(k)|^2\delta(k-k'),
$$
$$
\bigl(\psi_2^{(n)}(k),\psi_2^{(n)}(k')\bigr)=
2\pi |a_n(k)|^2\delta(k-k'), \quad k,\ k'>0,
$$
\be
\bigl(\psi_1^{(n)}(k),\psi_2^{(n)}(k')\bigr)=0,
\ee
are necessary for the further consideration.

Expending Green functions
\be
G^+_n(\rho,\rho')={1\over2\pi}\sum_{n=-\infty}^{\infty}
\e^{in(\theta-\theta')}G_n^{\pm}(\rho,\rho')
\ee
and using expressions (30), (34), (38), (39) one can find for partial
Green functions
\begin{eqnarray}
G^+_n(\rho,\rho')&=&{im\over a(k)k\hbar^2}
\bigl[\psi_1^{(n)}(\rho,k) \psi_2^{(n)}(\rho',k')\theta(\rho-
\rho')+\nonumber\\
&&+
\psi_2^{(n)}(\rho,k) \psi_1^{(n)}(\rho',k)\theta(\rho'-
\rho)\bigr],\nonumber\\
G^-_n(\rho,\rho')&=&(G^+_n(\rho,\rho'))^*
\end{eqnarray}

Substituting (31), (45), (46) in the Lippmann-Schwinger equations of
(29) and taking into account the expansion
\be
\psi^{(0)}_{\pm}(k)={\e^{ik\rho\cos(\theta-\theta_k)}\over2\pi}=
{\sqrt{\rho}\over2\pi}\sum_{n=-\infty}^{+\infty}
\e^{in(\theta-\theta_k)+{i\pi|n|\over2}}J_{|n|}(k\rho),
\ee
we get:
\begin{eqnarray}
\psi^{(+)}_{+,n}&=&{\psi_2^{(n)}\over2\pi a_n(k)}\sqrt{{1\over2\pi k}}\;
\e^{-in\theta_k+i\pi|n|+{i\pi\over4}},\\
\psi^{(-)}_{+,n}&=&{(\psi_2^{(n)})^*\over2\pi
a^*_n(k)}\sqrt{{1\over2\pi k}}\; \e^{-in\theta_k-{i\pi\over4}}, \\
\psi^{(-)}_{-,n}&=&{(\psi_1^{(n)})^*\over2\pi
a^*_n(k)}\sqrt{{1\over2\pi k}} \;\e^{-in\theta_k-{i\pi\over4}}.
\end{eqnarray}
Let us find, now, operators
$S_{++}$ and $S_{+-}$. They are completely defined by following matrix
elements:
\begin{eqnarray}
S_{++}(\vec{k},\vec{k'})&=&\bigl(\psi_+^{(0)}(\vec{k}),
S_{++}\psi_+^{(0)}(\vec{k'})\bigr)_{\cal{H}_+}=
\bigl(\psi_+^{(-)}(\vec{k}),\psi_+^{(+)}(\vec{k})\bigr)_{\cal{H}}
\nonumber\\
S_{+-}(\vec{k},\vec{k'})&=&\bigl(\psi_+^{(0)}(\vec{k}),
S_{+-}\psi_-^{(0)}(\vec{k'})\bigr)_{\cal{H}_+} =
\bigl(\psi_-^{(-)}(\vec{k}),\psi_+^{(+)}(\vec{k'})\bigr)_{\cal{H}}
\end{eqnarray}
Using expressions (31), (44), (48)--(50) one can find for matrix
elements (51):
\begin{eqnarray}
S_{++}(\vec{k},\vec{k'})&=&-{1\over2\pi\sqrt{kk'}}\e^{i\pi/2}\delta(k-k')
\sum_{n=-\infty}^{+\infty}\e^{in(\theta_k-\theta_{k'})+i\pi|n|}\;
{b_n^*(k)\over a_n(k)},
\end{eqnarray}
\begin{eqnarray}
S_{+-}(\vec{k},\vec{k'})&=&{1\over2\pi\sqrt{kk'}}\e^{i\pi/2}\delta(k-k')
\sum_{n=-\infty}^{+\infty}\e^{in(\theta_k-\theta_{k'})+i\pi|n|}\;
{1\over a_n(k)}.
\end{eqnarray}

The kernel of operator
$S_{++}$ (52) can be written in terms of the sum of singular and
regular, with respect to the angular variable, functions.
Actually, correspondingly to (2), (9), (13) and (35)
\begin{eqnarray}
\psi_1^{(n)}&=&\sqrt{{\pi
k(\rho+\rho_1)\over2}}\e^{i(-k\rho_1+{\pi\alpha_n^{(1)}\over2}+{\pi\over4})}
H^{(1)}_{\alpha_n^{(1)}}(k(\rho+\rho_1)),\quad \rho\geq P_1,\nonumber\\
\psi_2^{(n)}&=&\sqrt{{\pi
k(\rho+\rho_2)\over2}}\e^{i(k\rho_2+{\pi\alpha_n^{(2)}\over2}+{\pi\over4})}
H^{(1)}_{\alpha_n^{(2)}}(k|\rho+\rho_2|),\quad \rho\leq P_2
\end{eqnarray}
Using (39) we get from (54)
\begin{eqnarray}
\bar{\psi}_1^{(n)}&=&\sqrt{{\pi
k(\rho+\rho_1)\over2}}\e^{i(k\rho_1-{\pi\alpha_n^{(1)}\over2}-{\pi\over4})}
H^{(2)}_{\alpha_n^1}(k(\rho+\rho_1)),\nonumber\\
\bar{\psi}_2^{(n)}&=&\sqrt{{\pi
k(\rho+\rho_2)\over2}}\e^{i(-k\rho_2-{\pi\alpha_n^{(2)}\over2}-{\pi\over4})}
H^{(2)}_{\alpha_n^2}(k(rho+\rho_2)).
\end{eqnarray}
where
$$
P_i=\int\limits_{r_0}^{R_i}dr'\sqrt{f(r')},
$$
\be
\alpha_n^{(i)}=\biggl|{n-\Phi+{1\over2}\sigma^3\Phi_k^{(i)}\over1-
\Phi_k^{(i)}}
\biggr|,\qquad i=1,\ 2,
\ee
$$
\rho_i={R_i\over1-\Phi_n^{(i)}}-P_i.
$$
{}From (40) we find then:
\be
{b_n^*(k)\over a_n(k)}={W(\bar{\psi_n^{(1)}}(k),\psi_n^{(2)}(k))\over
W(\psi_n^{(1)}(k),\psi_n^{(2)}(k))}.
\ee

As is known, for large order and finite value of argument of Bessel
functions
\be
H_\nu^{(1)}(z)=-H_\nu^{(2)}(z)\biggl[1+O\biggl({ez\over2\nu}
\biggr)^{2\nu}\biggr].
\ee
Then, it follows from (54), (55), (57), (58) that
\be
b_n^*(k)=-\e^{2i(k\rho_1-{\pi\alpha_n^{(1)}\over2}-{\pi\over4})}a_n(k)
\biggl\{1+O\biggl[\biggl({ek(P_1+\rho_1)\over2\alpha_n^{(1)}}\biggr)^
{2\alpha_n^{(1)}}\biggr]\biggr\}
\ee
Using partial expression of unitarity (41) we find that
\be
{1\over |a_n|^2}=1-{|b_n|^2\over|a_n|^2}=1-\biggl|
1+O\biggl[\biggl({ek(P_1+\rho_1)\over2\alpha_n^{(1)}}\biggr)^
{2\alpha_n^{(1)}}\biggr]\biggr|^2=O\biggl(
{ek(P_1+\rho_1)\over2\alpha_n^{(1)}}\biggr)^{2\alpha_n^{(1)}}.
\ee
For small values of argument $(k(P_1+\rho_1))$ and
$\alpha_n^{(1)}\neq0$ the following expressions take place:
\be
H^{(2)}_{\alpha^{(1)}_n}\bigl(k(P_1+\rho_1)\bigr)=-H^{(1)}_{\alpha^{(1)}_n}
\bigl[1+O[(k(P_1+\rho_1))^{\alpha_n^{(1)}}]\bigr]
\ee
and for
$\alpha^{(1)}_n=0$:
\be
H^{(2)}_{0}\bigl((k(P_1+\rho_1)\bigr)=-H^{(1)}_{0}\bigl((k(P_1+\rho_1)\bigr)
\biggl[1+O\biggl({1\over|\ln(k(P_1+\rho_1))|}\biggr)\biggr],
\ee
\begin{eqnarray}
b_n^*(k)=-\e^{2i(k\rho_1-{\pi\alpha_n^{(1)}\over2}-{\pi\over4})}a_n(k)
\biggl\{1+O\biggl[\biggl(k(P_1+\rho_1)\biggr)^
{\alpha_n^{(1)}}\biggr]\biggr\}\;, \quad \alpha_n^{(1)}\neq 0,
\nonumber \\
b_n^*(k)=-\e^{2i(k\rho_1-{\pi\alpha_n^{(1)}\over2}-{\pi\over4})}a_n(k)
\biggl\{1+O\biggl[\biggl(\frac{1}{|\ln(k(P_1+\rho_1))|}\biggr)
\biggr]\biggr\}\;, \quad \alpha_n{(1)}=0.
\end{eqnarray}
\begin{eqnarray}
{1\over |a_n|^2}=1-{|b_n|^2\over|a_n|^2}=
O\biggl[\biggl(k(P_1+\rho_1)\biggr)^
{\alpha_n^{(1)}}\biggr]\;, \quad \alpha_n^{(1)}\neq 0,
\nonumber \\
{1\over |a_n|^2}=1-{|b_n|^2\over|a_n|^2}=
O\biggl[\biggl(\frac{1}{|\ln(k(P_1+\rho_1))|}\biggr)
\biggr]\;, \quad \alpha_n{(1)}=0.
\end{eqnarray}

It follows from (60) that series (53) for the kernel of the operator
$S_{+-}(\vec k,\vec k')$ converges uniformly with respect to the
angular variable and is therefore regular function of the angular
variable. Moreover, this function is infinitely differentiable
with respect to the angular variable and, in accordance with (64),
behaves like $O\left[\left(k(P_1+\rho_1)
\right)^{\alpha_n^{(1)}-1}\right]$ or $O\left[\left(k|\ln(k(P_1+
\rho_1))|\right)^{-1}\right]$ when $k(P_1+\rho_1) \to 0$.

Using (59) and (63) one can represent the kernel of the operator
$S_{++}(\vec k,\vec k')$ (52) in the following form:
\begin{eqnarray}
S_{++}(\vec{k},\vec{k}')&=&
{\delta(k-k')\over2\pi\sqrt{kk'}}\e^{2ik\rho_1}\sum_{n=-\infty}^{\infty}
\e^{in(\theta_k-\theta_{k'})+i\pi|n|-i\pi\alpha_n^{(1)}}-\nonumber\\
&-&{\delta(k-k')\over2\pi\sqrt{kk'}}\e^{2ik\rho_1}\sum_{n=-\infty}^{\infty}
\e^{in(\theta_k-\theta_{k'})+i\pi|n|-i\pi\alpha_n^{(1)}}\times\nonumber\\
&\times&\biggl[{b_n^*(k)\e^{i\pi/2+i\pi\alpha_n^{(1)}-2ik\rho_1}+a_n(k)
\over a_n(k)}\biggr].
\end{eqnarray}
The first sum in (65) is the kernel of $S$-matrix that describes
scattering of particle on coinciding singular gravitational and
magnetic strings. They are characterized by the curvature flux
$\Phi_K^{(1)}$ and magnetic flux
$\Phi$ respectively. We can write this sum as follows
\be
I'(k,\theta_k;k',\theta_{k'})+\delta(k-k')\sqrt{{i\over2\pi
k}}f(k;\theta_k-\theta_k')\e^{2ik\rho_1},
\ee
where $I'$ is modified unity operator. It is characterized by singular
kernel
$$
I'(k,\theta_k;k',\theta_{k'})={1\over2}{\delta(k-k')\over\sqrt{kk'}}
\e^{2ik\rho_1}\times
$$
\be
\times\biggl\{\Delta(\theta_k-\theta_{k'}+\beta_1\pi)
\exp[-i\mu_1(1+\beta_1)\pi]+\Delta(\theta_k-\theta_{k'}-\beta_1\pi)
\exp[i\mu_1(1+\beta_1)\pi]\biggr\}
\ee
where
$\Delta(\theta)={1\over2\pi}\sum_{n=-\infty}^{\infty}\e^{in\theta}$ --
periodic delta-function,
$\mu_1=\Phi-\sigma^3\Phi_k^{(1)}/2$;
$\beta_1=\Phi_K^{(1)}(1-\Phi_K^{(1)})^{-1}$;
$f$ -- scattering amplitude of particle on coinciding singular
gravitational and magnetic strings.

The second sum in (65) is regular infinitely differentiable function
with respect to the angular variable and its
behavior is the same as for
$S_{+-}(\vec{k},\vec{k'})$           for
$k(P_1+\rho_1)\to0$.
One can also formulate the problem of scattering of particle
directed from ``right'' to ``left'' cone.
For solving this problem it is necessary to define additionally
the operator
\be
\Omega^{(+)}_{-}(H,H_0^{(-)};J_-)=s-\lim_{t\to-\infty}\e^{iHt}J_-\e^{-
iH_0^{(-)}t}
\ee
or in terms of stationary scattering theory, wave function
\be
\psi_-^{(+)}=\Omega_-^{(+)}(H,H_0^{(-)};J_-)\psi^{(0)}_-.
\ee
It is easy to find it by the way same as above
\be
\psi_{-,n}^{(+)}={\psi^{(1)}_n\over2a_n(k)}\sqrt{{2\over\pi
k}}\e^{-in\theta_k+i\pi/4+i\pi|n|}.
\ee
In addition to the matrices of scattering
$S_{++}$ and $S_{+-}$ that describe probability of reflection to
``left'' cone and transmission to the ``right'' cone, respectively,
for particle
directed from left to right, we have to define matrices
$S_{--}$ and $S_{-+}$ describing reflection to the ``right'' cone and
transmission to ``left'' one:
\begin{eqnarray}
S_{--}&=&\bigl(\Omega_-^{(-)}(H,H^{(0)}_-;J_-)\bigr)^\dagger
\Omega_-^{(+)}(H,H^{(0)}_-;J_-);\nonumber\\
S_{-+}&=&\bigl(\Omega_+^{(-)}(H,H^{(0)}_+;J_-)\bigr)^\dagger
\Omega_-^{(+)}(H,H^{(0)}_-;J_-);
\end{eqnarray}
They are completely determined by their matrix elements:
\begin{eqnarray}
S_{--}(\vec{k},\vec{k}')&=&\bigl(\psi_-^{(0)}(\vec{k}),S_{--}
\psi_-^{(0)}(\vec{k}')\bigr)_{\cal{H}_-}=
\bigl(\psi_-^{(-)}(\vec{k}),
\psi_-^{(+)}(\vec{k}')\bigr)_{\cal{H}},
\nonumber\\
S_{-+}(\vec{k},\vec{k}')&=&\bigl(\psi_-^{(0)}(\vec{k}),S_{-+}
\psi_+^{(0)}(\vec{k}')\bigr)_{\cal{H}_-}=
\bigl(\psi_+^{(-)}(\vec{k}),
\psi_-^{(+)}(\vec{k}')\bigr)_{\cal{H}}.
\end{eqnarray}

Finally:
\begin{eqnarray}
S_{--}(\vec{k},\vec{k}')&=&
{\delta(k-k')\over2\pi\sqrt{kk'}}\e^{i\pi/2}\sum_{n=-\infty}^{\infty}
\e^{in(\theta_k-\theta_{k'})+i\pi|n|}\;{b_n(k)\over a_n(k)},
\nonumber\\
S_{-+}(\vec{k},\vec{k}')&=&
S_{+-}(\vec{k},\vec{k}')
\end{eqnarray}
It is possible to write the kernel of operator
$S_{--}(\vec{k},\vec{k}')$ in terms of the sum of singular and regular
functions with respect to the angular variable:
\begin{eqnarray}
S_{--}&=&
{\delta(k-k')\over2\pi\sqrt{kk'}}\sum_{n=-\infty}^{\infty}
\e^{in(\theta_k-\theta_{k'})+i\pi|n|-i\pi\alpha_n^{(2)}-2ik\rho}+
\nonumber\\
&+&{\delta(k-k')\over2\pi\sqrt{kk'}}\sum_{n=-\infty}^{\infty}
\e^{in(\theta_k-\theta_{k'})+i\pi|n|-i\pi\alpha_n^{(2)}-2ik\rho_2}\times
\nonumber\\
&\times&
\biggl[{b_n(k)\e^{i\pi\alpha_n^{(2)}+i\pi/2+2ik\rho_2}-a_n(k)\over
a_n(k)}\biggr].
\end{eqnarray}
{}From expressions (52)--(53), using the partial expression
of unitarity (41)
it is easy to get the following expression
\be (S_{++}S_{++}^\dagger)(\vec{k},\vec{k}')+
(S_{+-}S_{+-}^\dagger)(\vec{k},\vec{k}')={\delta(k-k')\over\sqrt{kk'}}
\Delta(\theta_k-\theta_{k'}).
\ee

Analogously for
$S_{--}$, $S_{-+}$
\be
(S_{--}S_{--}^\dagger)(\vec{k},\vec{k}')+
(S_{-+}S_{-+}^\dagger)(\vec{k},\vec{k}')={\delta(k-k')\over\sqrt{kk'}}
\Delta(\theta_k-\theta_{k'}).
\ee
The equations (75)-(76) are unitarity conditions for the problem under
consideration.

\vspace{.3in}
\noindent
{\bf Acknowledgments}\\
The work of A.V.M. and Yu.A.S. was supported by Swiss National Science
Foundation Grant No. CEEC/NIS/96-98/7 IP 051219.

\vfill
\eject

\end{document}